\documentclass[draft,
               nofootinbib,
               aps,
               prd,
               twocolumn,
               showpacs,
               superscriptaddress,
               preprintnumbers,
               floatfix]{revtex4-1}

\usepackage{dynlearn}
\usepackage{contour}

\colorlet{cd} {blue}
\colorlet{sd} {green!66!black}
\colorlet{id} {red}

\newcommand{\intrinsic}{\textcolor{id}{\textbf{\vphantom{gh}intrinsic}}\xspace}
\newcommand{\shared}{\textcolor{sd}{\textbf{\vphantom{gh}shared}}\xspace}
\newcommand{\synergistic}{\textcolor{cd}{\textbf{\vphantom{gh}synergistic}}\xspace}
\newcommand{\Intrinsic}{\textcolor{id}{\textbf{\vphantom{gh}Intrinsic}}\xspace}
\newcommand{\Shared}{\textcolor{sd}{\textbf{\vphantom{gh}Shared}}\xspace}
\newcommand{\Synergistic}{\textcolor{cd}{\textbf{\vphantom{gh}Synergistic}}\xspace}

\newcommand{\Xpast}{\ensuremath{X_{-1}}\xspace}
\newcommand{\Xpres}{\ensuremath{X_{0}}\xspace}
\newcommand{\Ypast}{\ensuremath{Y_{-1}}\xspace}
\newcommand{\Ypres}{\ensuremath{Y_{0}}\xspace}

\newcommand{\SKAR}{\ensuremath{S(X\!:\!Y~||~Z)}\xspace}
\renewcommand{\TDMI}{\I{\Xpast\!:\!\Ypres}\xspace}
\renewcommand{\TE}{\I{\Xpast \!:\!\Ypres\!\mid\!\Ypast}\xspace}
\renewcommand{\ITE}{\I{\Xpast\!:\!\Ypres \downarrow \Ypast}\xspace}
\newcommand{\corruption}{\ensuremath{\Pr(\overline{z}|z)}\xspace}

\begin{document}

\def\ourTitle{%
  Modes of Information Flow
}

\def\ourAbstract{%
Information flow between components of a system takes many forms and is key to
understanding the organization and functioning of large-scale, complex systems.
We demonstrate three modalities of information flow from time series
$X$ to time series $Y$. \emph{Intrinsic} information flow exists when the past
of $X$ is individually predictive of the present of $Y$, independent of $Y$'s
past; this is most commonly considered information flow. \emph{Shared}
information flow exists when $X$'s past is predictive of $Y$'s present in the
same manner as $Y$'s past; this occurs due to synchronization or common
driving, for example.  Finally, \emph{synergistic} information flow occurs when
neither $X$'s nor $Y$'s pasts are predictive of $Y$'s present on their own, but
taken together they are. The two most broadly-employed information-theoretic
methods of quantifying information flow---time-delayed mutual information and
transfer entropy---are both sensitive to a pair of these modalities:
time-delayed mutual information to both intrinsic and shared flow, and transfer
entropy to both intrinsic and synergistic flow. To quantify each mode
individually we introduce our \emph{cryptographic flow ansatz}, positing that
intrinsic flow is synonymous with secret key agreement between $X$ and $Y$.
Based on this, we employ an easily-computed secret-key-agreement
bound---intrinsic mutual information---to quantify the three flow modalities in
a variety of systems including asymmetric flows, coupled chaotic dynamical
systems, leader-follower collective dynamics, and financial markets.
}

\def\ourKeywords{%
  stochastic process, transfer entropy, intrinsic mutual information, mutual information, information flow.
}

\hypersetup{
  pdfauthor={James P. Crutchfield},
  pdftitle={\ourTitle},
  pdfsubject={\ourAbstract},
  pdfkeywords={\ourKeywords},
  pdfproducer={},
  pdfcreator={}
}

\author{Ryan G. James}
\email{rgjames@ucdavis.edu}
\affiliation{Complexity Sciences Center and Physics Department,
University of California at Davis, One Shields Avenue, Davis, CA 95616}

\author{Blanca Daniella Masante Ayala}
\email{masante@ucdavis.edu}
\affiliation{Complexity Sciences Center and Physics Department,
University of California at Davis, One Shields Avenue, Davis, CA 95616}

\author{Bahti Zakirov}
\email{bzakirov@ist.ac.at}
\affiliation{Institute of Science and Technology Austria,\\
Maria Gugging, Klosterneuburg, Austria}

\author{James P. Crutchfield}
\email{chaos@ucdavis.edu}
\affiliation{Complexity Sciences Center and Physics Department,
University of California at Davis, One Shields Avenue, Davis, CA 95616}

\date{\today}


\title{\ourTitle}

\begin{abstract}

\ourAbstract

\vspace{0.1in}

\end{abstract}


\preprint{\arxiv{1808.06723}}

\title{\ourTitle}

\date{\today}

\maketitle


\setstretch{1.1}

\listoffixmes

\section{Introduction}
\label{sec:introduction}

Information flow is an important signature of truly complex systems---an
incisive proxy for their structure and behavior. Truly complex systems are
becoming increasingly familiar to most all the sciences and engineering.
Certain expressed genes modulate the expression of others, leading to
structured biological processes in metabolism, development, and evolution. The
price of a good signals scarcity and demand to consumers and producers,
respectively. During specific cognitive tasks, only a small portion of the
brain's connectome is activated and so determining the underlying subnetwork is
critical to identifying neural function. To these ends, methods of tracking and
quantifying information flow form a core toolset for analyzing large-scale
complex systems.

Methodologically, it is difficult to find an alternative toolset with the many
features that recommend using information theory \cite{Cove06a,Mack03a} to
analyze complex systems. First, information accounts for any type of
co-relation \cite{Cove06a}. Whereas, statistical correlation requires an
underlying model and typically captures only linear statistical dependencies
\cite{Fell70a}. Second, information is broadly applicable. Many systems across
the sciences simply do not have an ``energy'' and this precludes appealing to
physics-based modeling that starts from a system Hamiltonian. In contrast,
information can be defined for mechanical, chemical, biological, social, and
engineered systems. Third, information provides directly comparable
quantitative units across qualitatively different systems. This is unlike
pairwise statistical correlation which comes in fixed, domain-specific
units---meters, concentration, volts, and dollars squared, and so on.
While information is universally measured in bits. Fourth, there is a rough
equivalence between probability theory and information theory \cite{Inga62a}
and a substantial foundational overlap between mathematical statistics and
information theory; e.g., see Refs. \cite{Kull68} and \cite[Ch. 11]{Cove06a}.

Finally, and perhaps most importantly, large-scale complex systems generate
emergent patterns---patterns that an analyst does not know a priori.
Information does not require prior knowledge of an appropriate representational
basis, which is essential when attempting to discover new patterns not seen
before \cite{Crut12a}. Despite a number of technical challenges, to date it
appears that the tools of information are the most general, workable
alternative available in the pursuit of complex systems. The following provides
new results that remove several of the remaining impediments to its even broader
use.

In particular, despite its long-lived intuitive appeal
\cite{Shan53a,Ashb57a,Shaw81,Kane86a,Horo14a}, information flow is not yet a
concretely defined concept. One consequence is that many methods of calculating
it fail surprisingly when deployed in unfamiliar contexts \cite{Jame15a}. Here,
we posit that these failures are not directly due to shortcomings of the
quantitative measures themselves, but rather arise due to analysts adhering to
the concept of a unitary information flow. In contrast, we demonstrate that the
flow between two time-series, $X = \ldots, X_t, X_{t+1}, \ldots$ and $Y =
\ldots, Y_t, Y_{t+1}, \ldots$, takes on three qualitatively
distinct modes: \emph{intrinsic}, \emph{shared}, and \emph{synergistic}
information flows. Unfortunately, to date methods of computing information flow
do not quantify these distinct modes. Here, we solve this problem via a novel
adaptation of cryptography.

Our development proceeds as follows. \Cref{sec:modes} discusses and exemplifies
the three modes of information flow. As an aid in this, App. \ref{app:Back}
briefly reviews the necessary notation and concepts from elementary probability
(random variables $X$ and $Y$), time series (random variable sequences
$X_{t:t'}$), and information measures ($\H{X}, \H{Y|X}, \I{X:Y}, \I{X:Y|Z}$).
\Cref{sec:measures} then surveys extant measures of information flow within
multivariate time series. \Cref{sec:cryptographic} introduces our
\emph{cryptographic flow ansatz} and shows how it isolates intrinsic
information flow and so yields the full three-way decomposition of information
flow. \Cref{sec:results} explores components of the decomposition in a variety
of settings, including asymmetric flows, coupled chaotic dynamical systems,
leader-follower dynamics in agent collectives, and financial indices.


A series of appendices rounds-out the main development, giving background on
information in time series and multivariate systems, differences with the
partial information decomposition, computational methods for information
measures, details on financial time series, and the statistical significance of
informational estimates.


\section{Modes of Information Flow}
\label{sec:modes}

Colloquially, information flow is the movement of information from one agent or
system to another. Unlike many flows considered in physics, such as electric
current or fluid flow, there is no single conservation law for information.
This makes quantifying information flow vastly more challenging. In light of
this, definitions of information flow have been somewhat ad-hoc, though they
typically employ either some form of mutual information~\cite{hlinka2018causal}
or quantify the influence one agent has on another~\cite{liang2013liang}. They
are often backed with examples where the proposed measure performs admirably,
though performance in other settings can be mixed or downright misleading. We
propose that these inconsistencies are due to conflating functionally distinct
modes of information flow.

Specifically, information flow from time series $X$ to time series $Y$ can take
three qualitatively distinct forms. The first, \emph{intrinsic flow}, is when
the past behavior of the $X$ time series is directly predictive of the present
behavior of the $Y$ time series in a fashion that the past behavior of $Y$ is
not. For example, this occurs when an ``upstream'' $X$ drives a ``downstream''
$Y$. The second mode, \emph{shared flow}, is when the present behavior of the
$Y$ time series can be inferred from the prior behavior of either the $X$ time
series or the $Y$ time series. This occurs due to, say, a common driver or
synchronization within a system. The third mode, \emph{synergistic flow},
occurs when both the past of the $X$ time series and the past of the $Y$ time
series are each independent of the present of the $Y$ time series, but when
combined the two pasts become predictive of it. This occurs in systems where
the behavior of a component strongly depends upon its context within the
system. Appendix \ref{app:joint} gives a more detailed rationale for the
three modalities of flow.

Markovian examples of these three types of flow are illustrated in
\cref{fig:timeseries}. Exemplifying intrinsic flow is the case where $X_{t}$ is
random and $Y$ simply follows it: $\Ypres = \Xpast$. Shared flow is
demonstrated with synchronization, where $\Ypres = \neg \Xpast = \neg \Ypast$,
where $\neg$ is the ``not'' operation. Finally, synergistic flow can be seen
when $X_{t}$ is random and $\Ypres = \Xpast \oplus \Ypast$, where $\oplus$ is
the exclusive-OR operation. Both shared and synergistic flows are symmetric, in
that they cannot be said to originate from either $X$'s past or $Y$'s past.
Whereas, intrinsic flow is uniquely attributed to $X$'s past.

Now, the question is: How do we detect and quantify these three forms of dependence from time series observations?

\begin{figure}
  \begin{tikzpicture}[every node/.style={transform shape}, scale=1.00]
    \begin{scope}
      \node at (0.5, 1.00) {\Intrinsic};
      \draw [very thick] (-0.25, 0.75) -- (1.25, 0.75);
      \node at (0, 0.5) {X};
      \node at (1, 0.5) {Y};
      \draw [very thick] (-0.25, 0.25) -- (1.25, 0.25);

      \node (nal) at (0, -0.00) {0};
      \node (nar) at (1, -0.00) {1};
      \node (nbl) at (0, -0.67) {0};
      \node (nbr) at (1, -0.67) {0};
      \node (ncl) at (0, -1.33) {1};
      \node (ncr) at (1, -1.33) {0};
      \node (ndl) at (0, -2.00) {0};
      \node (ndr) at (1, -2.00) {1};
      \node (nel) at (0, -2.67) {0};
      \node (ner) at (1, -2.67) {0};
      \node (nfl) at (0, -3.33) {0};
      \node (nfr) at (1, -3.33) {0};
      \node (ngl) at (0, -4.00) {1};
      \node (ngr) at (1, -4.00) {0};
      \node (nhl) at (0, -4.67) {0};
      \node (nhr) at (1, -4.67) {1};

      \draw [>=stealth, very thick, ->, color=id] (nal) -- (nbr);
      \draw [>=stealth, very thick, ->, color=id] (nbl) -- (ncr);
      \draw [>=stealth, very thick, ->, color=id] (ncl) -- (ndr);
      \draw [>=stealth, very thick, ->, color=id] (ndl) -- (ner);
      \draw [>=stealth, very thick, ->, color=id] (nel) -- (nfr);
      \draw [>=stealth, very thick, ->, color=id] (nfl) -- (ngr);
      \draw [>=stealth, very thick, ->, color=id] (ngl) -- (nhr);
    \end{scope}

    \begin{scope}[shift={(3.0, 0)}]
      \node at (0.5, 1.00) {\Shared};
      \draw [very thick] (-0.25, 0.75) -- (1.25, 0.75);
      \node at (0, 0.5) {X};
      \node at (1, 0.5) {Y};
      \draw [very thick] (-0.25, 0.25) -- (1.25, 0.25);

      \node (nal) at (0, -0.00) {0};
      \node (nar) at (1, -0.00) {0};
      \node (nbl) at (0, -0.67) {1};
      \node (nbr) at (1, -0.67) {1};
      \node (ncl) at (0, -1.33) {0};
      \node (ncr) at (1, -1.33) {0};
      \node (ndl) at (0, -2.00) {1};
      \node (ndr) at (1, -2.00) {1};
      \node (nel) at (0, -2.67) {0};
      \node (ner) at (1, -2.67) {0};
      \node (nfl) at (0, -3.33) {1};
      \node (nfr) at (1, -3.33) {1};
      \node (ngl) at (0, -4.00) {0};
      \node (ngr) at (1, -4.00) {0};
      \node (nhl) at (0, -4.67) {1};
      \node (nhr) at (1, -4.67) {1};

      \draw [>=stealth, very thick, ->, color=sd] (nal) -- (nbr);
      \draw [>=stealth, very thick, ->, color=sd] (nbl) -- (ncr);
      \draw [>=stealth, very thick, ->, color=sd] (ncl) -- (ndr);
      \draw [>=stealth, very thick, ->, color=sd] (ndl) -- (ner);
      \draw [>=stealth, very thick, ->, color=sd] (nel) -- (nfr);
      \draw [>=stealth, very thick, ->, color=sd] (nfl) -- (ngr);
      \draw [>=stealth, very thick, ->, color=sd] (ngl) -- (nhr);

      \draw [>=stealth, very thick, ->, color=sd] (nar.east) to [out=315, in=45] (nbr.north east);
      \draw [>=stealth, very thick, ->, color=sd] (nbr.east) to [out=315, in=45] (ncr.north east);
      \draw [>=stealth, very thick, ->, color=sd] (ncr.east) to [out=315, in=45] (ndr.north east);
      \draw [>=stealth, very thick, ->, color=sd] (ndr.east) to [out=315, in=45] (ner.north east);
      \draw [>=stealth, very thick, ->, color=sd] (ner.east) to [out=315, in=45] (nfr.north east);
      \draw [>=stealth, very thick, ->, color=sd] (nfr.east) to [out=315, in=45] (ngr.north east);
      \draw [>=stealth, very thick, ->, color=sd] (ngr.east) to [out=315, in=45] (nhr.north east);
    \end{scope}

    \begin{scope}[shift={(6.0, 0)}]
      \node at (0.5, 1.00) {\Synergistic};
      \draw [very thick] (-0.25, 0.75) -- (1.25, 0.75);
      \node at (0, 0.5) {X};
      \node at (1, 0.5) {Y};
      \draw [very thick] (-0.25, 0.25) -- (1.25, 0.25);

      \node (nal) at (0, -0.00) [outer sep=0] {0};
      \node (nar) at (1, -0.00) [outer sep=0] {1};
      \node (nbl) at (0, -0.67) [outer sep=0] {0};
      \node (nbr) at (1, -0.67) [outer sep=0] {1};
      \node (ncl) at (0, -1.33) [outer sep=0] {1};
      \node (ncr) at (1, -1.33) [outer sep=0] {1};
      \node (ndl) at (0, -2.00) [outer sep=0] {1};
      \node (ndr) at (1, -2.00) [outer sep=0] {0};
      \node (nel) at (0, -2.67) [outer sep=0] {0};
      \node (ner) at (1, -2.67) [outer sep=0] {1};
      \node (nfl) at (0, -3.33) [outer sep=0] {0};
      \node (nfr) at (1, -3.33) [outer sep=0] {1};
      \node (ngl) at (0, -4.00) [outer sep=0] {1};
      \node (ngr) at (1, -4.00) [outer sep=0] {1};
      \node (nhl) at (0, -4.67) [outer sep=0] {0};
      \node (nhr) at (1, -4.67) [outer sep=0] {0};

      \node (boxa) [draw, very thick, fit=(nal) (nar), inner sep=-0.5, rounded corners=0.1cm, color=cd] {};
      \node (boxb) [draw, very thick, fit=(nbl) (nbr), inner sep=-0.5, rounded corners=0.1cm, color=cd] {};
      \node (boxc) [draw, very thick, fit=(ncl) (ncr), inner sep=-0.5, rounded corners=0.1cm, color=cd] {};
      \node (boxd) [draw, very thick, fit=(ndl) (ndr), inner sep=-0.5, rounded corners=0.1cm, color=cd] {};
      \node (boxe) [draw, very thick, fit=(nel) (ner), inner sep=-0.5, rounded corners=0.1cm, color=cd] {};
      \node (boxf) [draw, very thick, fit=(nfl) (nfr), inner sep=-0.5, rounded corners=0.1cm, color=cd] {};
      \node (boxg) [draw, very thick, fit=(ngl) (ngr), inner sep=-0.5, rounded corners=0.1cm, color=cd] {};

      \draw [>=stealth, very thick, ->, color=cd] (boxa.east) to [out=315, in=45] (nbr.north east);
      \draw [>=stealth, very thick, ->, color=cd] (boxb.east) to [out=315, in=45] (ncr.north east);
      \draw [>=stealth, very thick, ->, color=cd] (boxc.east) to [out=315, in=45] (ndr.north east);
      \draw [>=stealth, very thick, ->, color=cd] (boxd.east) to [out=315, in=45] (ner.north east);
      \draw [>=stealth, very thick, ->, color=cd] (boxe.east) to [out=315, in=45] (nfr.north east);
      \draw [>=stealth, very thick, ->, color=cd] (boxf.east) to [out=315, in=45] (ngr.north east);
      \draw [>=stealth, very thick, ->, color=cd] (boxg.east) to [out=315, in=45] (nhr.north east);
    \end{scope}
  \end{tikzpicture}
  \caption{Modes of information flow:
    Intrinsic flow is exemplified by $\Ypres = \Xpast$.
    Shared flow by $\Ypres = \neg \Xpast = \neg \Ypast$.
    Synergistic flow by $\Ypres = \Xpast \oplus \Ypast$.
  }
  \label{fig:timeseries}
\end{figure}

\section{Extant Measures of Information Flow}
\label{sec:measures}

Historically, information flow has been measured via the \emph{time-delayed mutual information}~\cite{kantz2004nonlinear}:
\begin{align}
  \TDMI
  ~.
\end{align}
It posits that information flow from $X$ to $Y$ is the information shared
between $X$'s past observations \Xpast and $Y$'s present observation \Ypres.
As such, it is sensitive to both intrinsic and shared dependence, as seen in
\cref{fig:distributions}. While time-delayed mutual information captures a
restricted notion of causality, it ``\ldots fails to distinguish information
that is actually exchanged from shared information due to common history and
input signals '' \cite{schreiber2000measuring}. That is, it conflates intrinsic
and shared dependence.

To cleave away the shared dependence from the time-delayed mutual information
Ref.~\cite{schreiber2000measuring} proposed the \emph{transfer entropy}:
\begin{align}
  \TE
  ~,
\end{align}
---the information shared by $X$'s past and $Y$'s present, given $Y$'s
past---and correctly intuited that the influences of common history and input
signals on shared information ``\ldots~ are excluded by appropriate conditioning
of transition probabilities''. Unfortunately, this ignores the possibility of
conditional dependence. And, the transfer entropy suffers as a result: It fails
to distinguish intrinsic flow from synergistic flow; again, see
\cref{fig:distributions}. (That transfer entropy separates into two components
is not new, see App. \ref{app:pid}.) Taken at face value, this presents the
unfortunate situation of being unable to quantify any specific mode of
information flow.

A short aside will highlight the issue here. Previous efforts to measure
flow rest on the misunderstanding that conditioning is \emph{subtractive}: ``in
our new approach, these influences are excluded by appropriate conditioning of
transition probabilities'' \cite{schreiber2000measuring}. The erroneous
assumption here being that conditioning only \emph{excludes} dependency. In
point of fact, information-theoretic conditioning and probabilistic
conditioning, for that matter, are generically \emph{not} subtractive
operations, as demonstrated in \cref{fig:distributions} and Problem $10$ in
Chapter 2 of Ref. \cite{Cove91a}. Ignoring this has led to the belief that
conditioning on more and more time series results in a more incisive analysis
of information flow within a
system~\cite{schreiber2000measuring,sun2014causation,bossomaierintroduction}.
It need not.

\begin{figure}
  \centering
  \begin{tikzpicture}[every node/.style={transform shape}, scale=0.855]
    \node (distcd) at (-0.3, 0) [anchor=west] {
      \begin{tabular}{cccc}
        \multicolumn{4}{c}{\large\textcolor{id}{Intrinsic}} \\
        \toprule
        \Xpast & \Ypres & \Ypast & $\Pr$ \\
        \midrule
        \textcolor{id}{0} & \textcolor{id}{0} & \textcolor{id}{0} & $\nicefrac{1}{4}$ \\
        \textcolor{id}{0} & \textcolor{id}{0} & \textcolor{id}{1} & $\nicefrac{1}{4}$ \\
        \textcolor{id}{1} & \textcolor{id}{1} & \textcolor{id}{0} & $\nicefrac{1}{4}$ \\
        \textcolor{id}{1} & \textcolor{id}{1} & \textcolor{id}{1} & $\nicefrac{1}{4}$ \\
        \bottomrule
      \end{tabular}
    };
    \node (distsd) at ($ (distcd.north east) + (0.4, 0) $) [anchor=north west] {
      \begin{tabular}{cccc}
        \multicolumn{4}{c}{\large\textcolor{sd}{Shared}} \\
        \toprule
        \Xpast & \Ypres & \Ypast & $\Pr$ \\
        \midrule
        \textcolor{sd}{0} & \textcolor{sd}{1} & \textcolor{sd}{0} & $\nicefrac{1}{2}$ \\
        \textcolor{sd}{1} & \textcolor{sd}{0} & \textcolor{sd}{1} & $\nicefrac{1}{2}$ \\
        \bottomrule
      \end{tabular}
    };
    \node (distid) at ($ (distsd.north east) + (0.4, 0) $) [anchor=north west] {
      \begin{tabular}{cccc}
        \multicolumn{4}{c}{\large\textcolor{cd}{Synergistic}} \\
        \toprule
        \Xpast & \Ypres & \Ypast & $\Pr$ \\
        \midrule
        \textcolor{cd}{0} & \textcolor{cd}{0} & \textcolor{cd}{0} & $\nicefrac{1}{4}$ \\
        \textcolor{cd}{0} & \textcolor{cd}{1} & \textcolor{cd}{1} & $\nicefrac{1}{4}$ \\
        \textcolor{cd}{1} & \textcolor{cd}{0} & \textcolor{cd}{1} & $\nicefrac{1}{4}$ \\
        \textcolor{cd}{1} & \textcolor{cd}{1} & \textcolor{cd}{0} & $\nicefrac{1}{4}$ \\
        \bottomrule
      \end{tabular}
    };

    \node (eqscd) at ($(distcd.north) + (0, -3)$) [anchor=north] {
      \begin{minipage}{0.4\columnwidth}
        \small
        \begin{align*}
          \TDMI &= \textcolor{id}{1} \\
          \TE &= \textcolor{id}{1}
        \end{align*}
      \end{minipage}
    };
    \node (eqssd) at ($(distsd.north) + (0, -3)$) [anchor=north] {
      \begin{minipage}{0.4\columnwidth}
        \small
        \begin{align*}
          \TDMI &= \textcolor{sd}{1} \\
          \TE &= \textcolor{sd}{0}
        \end{align*}
      \end{minipage}
    };
    \node (eqsid) at ($(distid.north) + (0, -3)$) [anchor=north] {
      \begin{minipage}{0.4\columnwidth}
        \small
        \begin{align*}
          \TDMI &= \textcolor{cd}{0} \\
          \TE &= \textcolor{cd}{1}
        \end{align*}
      \end{minipage}
    };
  \end{tikzpicture}
\caption{Three canonical types of dependence between two variables
	\Xpast and \Ypres in the context of a third \Ypast:
    (Left) Intrinsic dependence exists between the first two variables in spite
	of the third.
    (Middle) Shared dependence exists synchronously with the third.
    (Right) Synergistic dependence exists only when also observing the third.
	}
\label{fig:distributions}
\end{figure}

\section{Cryptographic Common Information}
\label{sec:cryptographic}

Overcoming the challenge of information flow requires adopting a different
viewpoint---one that directly addresses when two system components, and only
two system components, possess common information. Solving this would
circumvent the open-ended issue of conditioning on all of a system's other,
possibly unspecified, variables besides the two of interest; a strategy that,
on its own, fatally ignores conditional dependence, as we just argued. Thus, we
must simultaneously solve a definitional problem and a technical problem:
respectively (i) acknowledging distinct modes of flow and (ii) accounting for
both conditional independence and conditional dependence. Our solution appeals
to cryptography and the information theory of two parties sharing secret keys.
We introduce an ansatz that directly quantifies intrinsic flow and, thereby,
completes the decomposition of information flow into its three modes.

Consider again the flow of intrinsic information from \Xpast to \Ypres. The
flow implicates some sort of dependency or correlation between $X$ and $Y$ that
can unambiguously be attributed to $X$. Were this dependency able to be
reproduced from other aspects of the system, it could not be said to have
originated from \Xpast. This observation evokes the cryptographic idea of
\emph{secret key agreement} and leads to our ansatz:
\begin{tcolorbox}[title=Cryptographic Flow Ansatz]
Intrinsic information flow exists exactly when \Xpast and \Ypres can agree on
a secret key, while the past of the rest of the system eavesdrops.
\end{tcolorbox}
Quantitatively, intrinsic information flow is the rate at which information
secret to \Xpast and \Ypres can be extracted from observations of the system.

\subsection{Secret Key Agreement Rate}
\label{subsec:skar}

In this way, we identify the intrinsic information flow as the secret-key
agreement rate~\cite{maurer1993secret}, which is defined as follows. Consider a
joint random variable $(X, Y, Z)$, where Alice has access to the $X$
realizations, Bob the $Y$ realizations, and Eve the $Z$ realizations. Given $N$
IID realizations of the joint variable, let $X^N$ denote Alice's observations,
$Y^N$ Bob's, and $Z^N$ Eve's. Random variable $V$ represents the public
communication that all three observe. Let $S$ denote the secret key that Alice
and Bob wish to have in common.

Now, let $S_X$ and $S_Y$ represent the secret keys that Alice and Bob,
respectively, distill from their private observations as well as from the
public communication $V$: $S_X = f(X^N, V)$ and $S_Y = g(Y^N, V)$. In this,
functions $f$ and $g$ represent the mechanism by which Alice and Bob construct
their copy of secret key $S$. A \emph{secret key agreement scheme} defines the
allowed public communications $V$ and mechanisms $f$ and $g$. If the scheme is
any good, $S_X$ and $S_Y$ will be identical and equal to $S$ with high
probability: $\Pr(S_X = S_Y = S) \geq 1 - \epsilon$. Moreover, being secret,
the key $S$ should have arbitrarily small correlation with Eve's private
observations $Z$ and the public communication $V$: $\I{S : V, Z^N} \leq
\epsilon$. The \emph{secret-key agreement rate} \SKAR then is the maximum rate
$R$ such that:
\begin{align*}
  \lim_{N \to \infty} \frac{1}{N} \H{S} \geq R - \epsilon
  ~,
\end{align*}
for $N > 0$ and $\epsilon > 0$.
In other words, the secret key agreement rate is the largest rate at which a
secret key $S$ can be successfully produced.

Effectively, given realizations of the three-way joint random variable, there
exists a scheme by which Alice and Bob can publicly exchange information and
then distill their public and private information into a secret key upon which
they both agree with arbitrarily high probability, but which has arbitrarily
little information shared with all information available to Eve. The challenge
now is to determine the secret-key extraction functions $f$ and $g$, as well as
what public communication $V$ is necessary. See Ref.~\cite{jostinformation} for
concrete examples.

\subsection{An Easily Computed Upper Bound}
\label{subsec:upper_bound}

Though we identified the secret key agreement rate with intrinsic information
flow, the nonconstructive nature of its definition mandates we appeal to some
proxy if we wish to practically estimate it. While several lower and upper
bounds exist for the secret key agreement rate, here we use the \emph{intrinsic
mutual information}~\cite{maurer1999unconditionally}. We recommend this upper
bound due to its nontrivial behavior (exemplified shortly) and relatively
straightforward estimation.

An eavesdropper not only has access to her observations $z$, but also to any
(local) modification \corruption of them. Therefore, the ability of Alice and
Bob to agree upon a key cannot be reduced if $Z$ is replaced by any
``corruption'' $\overline{Z}$. This observation simplifies the secret key
optimization, leading to a constructive bound on the secret key agreement rate:
\begin{align}
  \SKAR &\leq \min_{\corruption} \I{X : Y \mid \overline{Z}} \\
        &= \I{X : Y \downarrow Z}
  ~.
\end{align}
The last quantity---\emph{intrinsic mutual information}---is therefore an upper
bound on the secret key agreement rate. It can be easily verified that \I{X : Y
\downarrow Z} is bounded from above by both \I{X : Y}, when \corruption is
constant, and \I{X : Y \mid Z}, when \corruption is the identity. Fortunately,
this optimization is not difficult due to the boundedness of $\overline{Z}$.
(In fact, $|\overline{Z}| \leq |Z|$~\cite{christandl2003property}).
Appendix \ref{app:computing} explains how to calculate the intrinsic mutual
information.

To illustrate its behavior, consider the distribution in
\cref{fig:intrinsicdist}~\cite{maurer1999unconditionally}. This distribution
has two qualitatively distinct sets of events. The first, encoded using 0s and
1s, exhibits conditional dependence: any pair of $X$, $Y$, or $Z$ are
independent, but given the third they are perfectly correlated. The second,
encoded using 2s and 3s, exhibits conditional independence: any pair is
perfectly correlated and is also correlated with the third. The mutual
information $\I{X : Y} = \H{\nicefrac{1}{2}, \nicefrac{1}{4}, \nicefrac{1}{4}}
= \SI{\nicefrac{3}{2}}{\bit}$, reflecting that they share 01-, 2-, and 3-ness. However, the
conditional mutual information $\I{X : Y \mid Z} = \SI{\nicefrac{1}{2}}{\bit}$, reflecting
the conditional dependence that occurs half the time. The mapping of $Z$ to
$\overline{Z}$ given in the table demonstrates that the conditional dependence
can be destroyed while preserving the conditional independence: $\I{X : Y \mid
\overline{Z}} = \SI{0}{\bit} = \I{X : Y \downarrow Z}$. This indicates that
both the mutual information and conditional mutual information misleadingly
identify dependencies that do not belong to $X$ and $Y$ alone, but rather are
shared by, or induced by, $Z$.

\begin{figure}
  \begin{tikzpicture}
    \node (table) at (0, 0) [anchor=north east] {
      \begin{tabular}{ccccc}
        \toprule
        $X$ & $Y$ & $Z$ & $\overline{Z}$ & $\Pr$ \\
        \midrule
        0 & 0 & 0 & 0 & $\nicefrac{1}{8}$ \\
        0 & 1 & 1 & 0 & $\nicefrac{1}{8}$ \\
        1 & 0 & 1 & 0 & $\nicefrac{1}{8}$ \\
        1 & 1 & 0 & 0 & $\nicefrac{1}{8}$ \\
        2 & 2 & 2 & 2 & $\nicefrac{1}{4}$ \\
        3 & 3 & 3 & 3 & $\nicefrac{1}{4}$ \\
        \bottomrule
      \end{tabular}
    };
    \node (eq1) at (4, -1) [anchor=east] {$\I{X : Y} = \SI{\nicefrac{3}{2}}{\bit}$};
    \node (eq2) at (4, -2) [anchor=east] {$\I{X : Y \mid Z} = \SI{\nicefrac{1}{2}}{\bit}$};
    \node (eq3) at (4, -3) [anchor=east] {$\I{X : Y \downarrow Z} = \SI{0}{\bit}$};
  \end{tikzpicture}
\caption{The intrinsic mutual information \I{X : Y \downarrow Z} can be less
	than both \I{X : Y} and \I{X : Y \mid Z}: The mutual information \I{X : Y}
	captures the $01$-, $2$-, and $3$-ness that is shared by $X$ and $Y$, while
	the conditional mutual information \I{X : Y \mid Z} captures the fact that
	knowledge of $Z$ indicates whether the $0$s and $1$s of $X$ and $Y$ are the
	same or different. Neither of these dependencies are held by $X$ and $Y$
	alone, and so the intrinsic mutual information \I{X : Y \downarrow Z}
	vanishes.
  }
  \label{fig:intrinsicdist}
\end{figure}


\subsection{Flow Decomposition}
\label{subsec:decomposition}

Recall that we showed the time-delayed mutual information \TDMI captures both
intrinsic and shared flows, while transfer entropy \TE captures intrinsic and
synergistic flows. Together with our cryptographic flow ansatz that quantifies
intrinsic flow, simple algebra now gives a full and constructive decomposition
of the distinct flow modes:
\begin{description}
\setlength{\topsep}{0pt}
\setlength{\itemsep}{0pt}
\setlength{\parsep}{0pt}
  \item[Intrinsic Flow] \ITE
  \item[Shared Flow] $\TDMI~-~\ITE$
  \item[Synergistic Flow] $\TE~-~\ITE$
\end{description}


To illustrate how this works, \cref{fig:distributions_b} returns to the
canonical dependency types of \cref{sec:modes} and applies the decomposition to
each. In the case of the Intrinsic pair of time series, we find that intrinsic
flow is \SI{1.0}{\bit} while shared and synergistic are both \SI{0.0}{\bit}.
The synchronized or Shared pair of time series decomposes with \SI{1.0}{\bit}
of shared flow, with \SI{0.0}{\bit} of both intrinsic and synergistic flows.
Finally, the Synergistic pair has no intrinsic or shared information flow, but
\SI{1.0}{\bit} of synergistic flow. These observations justify the names given
to the three modes.

\begin{figure}
  \centering
  \begin{tikzpicture}[every node/.style={transform shape}, scale=0.805]
    \node (distcd) at (0, 0) [anchor=west] {
      \begin{tabular}{cccc}
        \multicolumn{4}{c}{\large\Intrinsic} \\
        \toprule
        \Xpast & \Ypres & \Ypast & $\Pr$ \\
        \midrule
        \textcolor{id}{0} & \textcolor{id}{0} & \textcolor{id}{0} & $\nicefrac{1}{4}$ \\
        \textcolor{id}{0} & \textcolor{id}{0} & \textcolor{id}{1} & $\nicefrac{1}{4}$ \\
        \textcolor{id}{1} & \textcolor{id}{1} & \textcolor{id}{0} & $\nicefrac{1}{4}$ \\
        \textcolor{id}{1} & \textcolor{id}{1} & \textcolor{id}{1} & $\nicefrac{1}{4}$ \\
        \bottomrule
      \end{tabular}
    };
    \node (distsd) at ($ (distcd.north east) + (0.7, 0) $) [anchor=north west] {
      \begin{tabular}{cccc}
        \multicolumn{4}{c}{\large\Shared} \\
        \toprule
        \Xpast & \Ypres & \Ypast & $\Pr$ \\
        \midrule
        \textcolor{sd}{0} & \textcolor{sd}{1} & \textcolor{sd}{0} & $\nicefrac{1}{2}$ \\
        \textcolor{sd}{1} & \textcolor{sd}{0} & \textcolor{sd}{1} & $\nicefrac{1}{2}$ \\
        \bottomrule
      \end{tabular}
    };
    \node (distid) at ($ (distsd.north east) + (0.7, 0) $) [anchor=north west] {
      \begin{tabular}{cccc}
        \multicolumn{4}{c}{\large\Synergistic} \\
        \toprule
        \Xpast & \Ypres & \Ypast & $\Pr$ \\
        \midrule
        \textcolor{cd}{0} & \textcolor{cd}{0} & \textcolor{cd}{0} & $\nicefrac{1}{4}$ \\
        \textcolor{cd}{0} & \textcolor{cd}{1} & \textcolor{cd}{1} & $\nicefrac{1}{4}$ \\
        \textcolor{cd}{1} & \textcolor{cd}{0} & \textcolor{cd}{1} & $\nicefrac{1}{4}$ \\
        \textcolor{cd}{1} & \textcolor{cd}{1} & \textcolor{cd}{0} & $\nicefrac{1}{4}$ \\
        \bottomrule
      \end{tabular}
    };

    \node (eqscd) at ($(distcd.north) + (0, -3)$) [anchor=north] {
      \begin{minipage}{0.4\columnwidth}
        \small
        \begin{align*}
          \intrinsic   &= \textcolor{id}{1} \\
          \shared      &= \textcolor{sd}{0} \\
          \synergistic &= \textcolor{cd}{0}
        \end{align*}
      \end{minipage}
    };
    \node (eqssd) at ($(distsd.north) + (0, -3)$) [anchor=north] {
      \begin{minipage}{0.4\columnwidth}
        \small
        \begin{align*}
          \intrinsic   &= \textcolor{id}{0} \\
          \shared      &= \textcolor{sd}{1} \\
          \synergistic &= \textcolor{cd}{0}
        \end{align*}
      \end{minipage}
    };
    \node (eqsid) at ($(distid.north) + (0, -3)$) [anchor=north] {
      \begin{minipage}{0.4\columnwidth}
        \small
        \begin{align*}
          \intrinsic   &= \textcolor{id}{0} \\
          \shared      &= \textcolor{sd}{0} \\
          \synergistic &= \textcolor{cd}{1}
        \end{align*}
      \end{minipage}
    };

  \end{tikzpicture}
\caption{Flow decomposition of the canonical dependency types of
	\cref{fig:distributions}: Each dependency type is associated with a unique
	flow mode: intrinsic with intrinsic, shared with shared, and synergistic
	with synergistic.
  }
  \label{fig:distributions_b}
\end{figure}

\section{Applications}
\label{sec:results}

With their definitions and estimation methods laid out, we now turn to
demonstrate the diversity and advantage of quantifying separate modes of
information flow in settings ranging from
asymmetric flow, coupled chaotic dynamical systems, and leader-follower
group dynamics to financial indices. We do end, though, with a
cautionary exploration of biases that arise in estimating nonnegative
statistics, such as the information measures here.

\subsection{Asymmetric Information Flow?}
\label{subsec:example}

At first blush, it seems difficult for there to be shared information flow from
$X$ to $Y$, but not from $Y$ to $X$. Here, we provide a relatively simple
example and an intuitive explanation of this phenomenon.

Consider the following (jointly) Markovian transition matrix $T$ defining a
pair of time series $X$ and $Y$:
\begin{align}
  T_{\Xpast, \Ypast \rightarrow \Xpres, \Ypres} = \bordermatrix{
             & 0, 0              & 0, 1              & 1, 0 \cr
        0, 0 & 0                 & 1                 & 0    \cr
        0, 1 & 0                 & 0                 & 1    \cr
        1, 0 & $\nicefrac{4}{9}$ & $\nicefrac{5}{9}$ & 0
      }
  ~,
\end{align}
with stationary distribution
$\pi = \left( \nicefrac{2}{11}, \nicefrac{9}{22}, \nicefrac{9}{22} \right)$.
The time series' temporal joint distribution is:
\begin{table}[h!]
  \begin{tabular}{ccccc}
    \toprule
    \Xpast & \Ypast & \Xpres & \Ypres & $\Pr$ \\
    \midrule
    0 & 0 & 0 & 1 & \nicefrac{2}{11} \\
    0 & 1 & 1 & 0 & \nicefrac{9}{22} \\
    1 & 0 & 0 & 0 & \nicefrac{2}{11} \\
    1 & 0 & 0 & 1 & \nicefrac{5}{22} \\
    \bottomrule
  \end{tabular}
\end{table}

The joint time series has information flows from $Y$ to $X$ of
\SI{0.526200}{\bit} intrinsic, \SI{0.449821}{\bit} shared, and \SI{0.0}{\bit}
synergistic; and from $X$ to $Y$ of \SI{0.0}{\bit} intrinsic,
\SI{0.044381}{\bit} shared, and \SI{0.120759}{\bit} synergistic. Note that from
$Y$ to $X$, the intrinsic information flow is equal to the transfer entropy,
while from $Y$ to $X$ the intrinsic information flow is equal to the
time-delayed mutual information.

To gain a more intuitive understanding of the asymmetry in this, let us isolate
the transition matrix corresponding to the generation of the $X$ time series:
\begin{align}
  T\left[\Xpast, \Ypast \rightarrow \Xpres\right] = \bordermatrix{
             & 0 & 1 \cr
        0, 0 & 1 & 0 \cr
        0, 1 & 0 & 1 \cr
        1, 0 & 1 & 0
      }
  ~.
\end{align}
In this instance, $\I{\Xpast : \Ypast} = \SI{0.449821}{\bit}$ while $\I{\Xpast
: \Ypast | \Xpres} = \SI{0.0}{\bit}$. This indicates that the variables form a
Markov chain $\Xpast - \Xpres - \Ypast$ and so any information \Xpast and
\Ypast share is contained within \Xpres. Therefore, the shared information flow
is $\SI{0.449821}{\bit}$. From this, the intrinsic information flow from $Y$ to
$X$ is $\I{\Ypast : \Xpres} - \SI{0.449821}{\bit} = \SI{0.526200}{\bit}$ and
synergistic is $\I{\Ypast : \Xpres | \Xpast} - \SI{0.449821}{\bit} =
\SI{0.0}{\bit}$.

Looking toward the $Y$ time series, we find the following transition matrix:
\begin{align}
  T\left[\Xpast, \Ypast \rightarrow \Ypres\right] = \bordermatrix{
             & 0                 & 1                 \cr
        0, 0 & 0                 & 1                 \cr
        0, 1 & 1                 & 0                 \cr
        1, 0 & $\nicefrac{4}{9}$ & $\nicefrac{5}{9}$
      }
  ~.
\end{align}
Here, consider a locally-modified $\Ypast^\prime$ constructed by passing \Ypast
through a channel that preserves the value $0$, but maps a $1$ to a $0$ with
probability $\nicefrac{16}{45}$. This results in the modified transition matrix:
\begin{align}
  T\left[\Xpast, \Ypast^\prime \rightarrow \Ypres\right] = \bordermatrix{
             & 0                 & 1                 \cr
        0, 0 & $\nicefrac{4}{9}$ & $\nicefrac{5}{9}$ \cr
        0, 1 & 1                 & 0                 \cr
        1, 0 & $\nicefrac{4}{9}$ & $\nicefrac{5}{9}$
      }
  ~,
\end{align}
given $\Ypast^\prime$, \Xpast, and \Ypres are independent.

This implies that no information about \Ypres can be uniquely attributed to \Xpast, since there is a method of reconstructing any influence \Xpast has on \Ypres using \Ypast alone. We then conclude that the intrinsic information flow from $X$ to $Y$ is \SI{0.0}{\bit}.
Shared information flow is then $\I{\Xpast : \Ypres} = \SI{0.044381}{\bit}$, while synergistic flow is $\I{\Xpast : \Ypres | \Ypast} = \SI{0.120759}{\bit}$.

These jointly Markovian time series exemplify the degree of asymmetry that can
exist in information flow. Specifically, it is not immediately obvious that
shared information flow---due to common driving or synchronization, for
example---can be large in one direction while small or nonexistent in the
other. These sorts of asymmetries appear in a variety of data sets, and so its
demonstration in a relatively simple Markovian setting is pedagogically
helpful and must be kept in mind when structurally interpreting the
estimated flows. These relationships are summarized in \cref{fig:markov}.

\begin{figure}
  \centering
  \begin{tikzpicture}[every node/.style={transform shape}, scale=1.0, style=vaucanson]
    \node [state] (x) at (180:2.5cm)  {$x$};
    \node [state] (y) at (0:2.5cm)    {$y$};

    \draw [>=stealth, ->, color=sd, line width=0.176mm] (x) to [out=45, in=135] (y);
    \draw [>=stealth, ->, color=cd, line width=0.484mm] (x) to [out=25, in=155] (y);

    \draw [>=stealth, ->, color=id, line width=2.104mm] (y) to [out=205, in=335] (x);
    \draw [>=stealth, ->, color=sd, line width=1.796mm] (y) to [out=225, in=315] (x);
  \end{tikzpicture}
\caption{Information flows between $x$ and $y$: The arrow color corresponds to
	flow mode: Red for intrinsic, green for shared, and blue for synergistic.
	Arrow width corresponds to the flow strength: Wider the arrow, the larger
	that flow mode. For the case considered in this section, there is
	relatively little flow from $x$ to $y$ and its modes are shared and
	synergistic, while there is significantly more information flow from $y$ to
	$x$ and its modes are intrinsic and shared.
  }
\label{fig:markov}
\end{figure}

\subsection{Coupled Chaotic Maps}
\label{subsec:coupledtentmap}

Shifting from the setting of joint stochastic processes, consider now
a pair of coupled deterministic, but chaotic dynamical systems---coupled tent
maps of the unit square. Individually, a tent map is a simple discrete-time
dynamical system that exhibits chaotic behavior. The update rule is given by:
\begin{align*}
  x_{t+1} &= f(a, x_t) \\
  f(a, x) &= a \left( \frac{1}{2} - \left|x - \frac{1}{2}\right| \right)
  ~,
\end{align*}
where the initial condition $x_0 \in [0,1]$. A tent map is chaotic for $1 < a
\leq 2$. An \emph{orbit} of the map---a temporal sequence $x_0, x_1,
\ldots$---can be discretized via the mapping:
\begin{align*}
  s_t = \begin{cases}
    0 & \text{if } x_t < \frac{1}{2} \\
    1 & \text{otherwise}
  \end{cases}
  ~.
\end{align*}
This coarse binary discretization---mapping from a real number to a binary
one---is known as a \emph{generating partition} of the unit interval.
Importantly, the resulting binary time series $s_0, s_1, \ldots$---preserves
many dynamical properties of the continuous-space orbit, such as the degree of
amplifying small variations in the state, captured by the Kolmogorov-Sinai
entropy rate \cite{Laso85a}.

\begin{figure}
  \centering
  \includegraphics{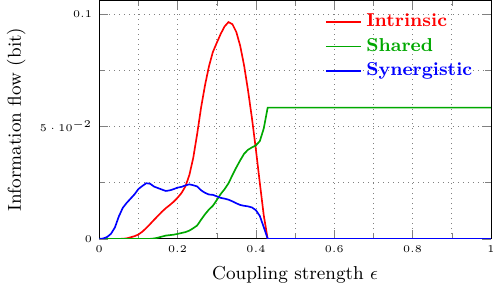}
  \caption{\Intrinsic, \shared, and \synergistic information flows from $A$ to
	$B$ for the coupled tent maps with $a_1 = a_2 = 1.75$, as a function of
	the coupling strength $\epsilon$. Beyond $\epsilon \approx 0.43$ the two
	maps synchronize, the entire flow is shared, and the time series from each
	map are the same.
	}
  \label{fig:coupledtentmaps}
\end{figure}

We couple two such systems together via the mapping:
\begin{align*}
  x_{t+1} &= f(a_1, x_t) \\
  y_{t+1} &= f(a_2, \epsilon x_t + (1-\epsilon) y_t) 
    ~,
\end{align*}
where $\epsilon \in [0,1]$ is a coupling strength. Furthermore, we discretize
the two orbits individually via the same mapping as above, resulting in the
discrete-time binary sequences $A_t$ and $B_t$. We can then estimate the
intrinsic, shared, and synergistic information flows from the ``source'' map $x_t$ to the ``target'' map $y_t$,
shown in \cref{fig:coupledtentmaps}, from the sequence statistics in the time
series $s_0, s_1, \ldots$ ($A_t$ and $B_t$ respectively) from each tent map.

\begin{figure}
  \centering
  \includegraphics{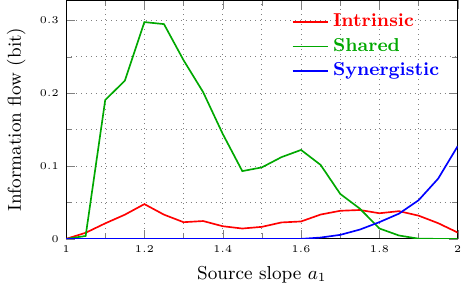}
  \caption{\Intrinsic, \shared, and \synergistic information flows from $A$ to
	$B$ for the coupled tent maps at coupling $\epsilon = 0.40$ with the
	target slope fixed at $a_2 = 1.75$, as a function of the source slope
	$a_1$.
	}
  \label{fig:tentmapslopeone}
\end{figure}

\begin{figure}
  \centering
  \includegraphics{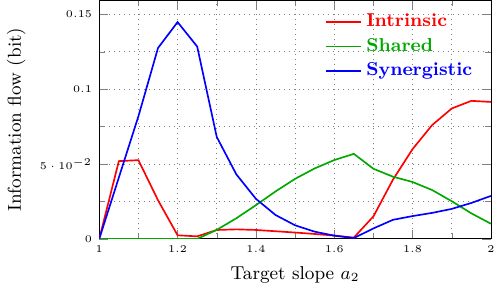}
  \caption{\Intrinsic, \shared, and \synergistic information flows from $A$ to
	$B$ for the coupled tent maps at coupling $\epsilon = 0.40$ with the
	source slope fixed at $a_1 = 1.75$, as a function of the target slope
	$a_2$.
	}
  \label{fig:tentmapslopetwo}
\end{figure}

Alternatively, we can hold the coupling fixed at $\epsilon = 0.40$ and vary the
slope of either map. \Cref{fig:tentmapslopeone} varies the source slope $a_1$
while holding $a_2 = 1.75$, and \cref{fig:tentmapslopetwo} varies the target
slope $a_2$ while holding $a_1 = 1.75$.

Together these estimates of the informational modes demonstrate the range of
possible information flows between the coupled chaotic dynamical systems.

In particular, \Cref{fig:coupledtentmaps} demonstrates that for strongly
coupled maps---above coupling $\epsilon \approx 0.43$---the intrinsic and
synergistic informations flows vanish and the shared information is maximized.
This occurs since the states of the component maps become synchronized and so
the binary time series from each is the same---all the available (chaotically
generated) information is shared, with none arising from intrinsic or
synergistic interactions.

Below that coupling strength one finds a rich variation in the maps' dynamic
behaviors. There is a peak in intrinsic flow around $\epsilon \approx 0.33$
where the maps are individually most chaotic and generate the most information
(high entropy rate), and that information is common between them. For $\epsilon
\in [0, 0.43]$ there is a cooperative interaction between the map's dynamics that
leads to a moderate amount of synergistic information flow being created.

In contrast, \Cref{fig:tentmapslopeone} illustrates varying the nonlinearity
parameter $a_1$ for the first map and this increases the dynamical instability
there. This appears then to be shared with the second chaotic map (at $a_2 =
1.75$). The first map's chaotic behavior appears to dominate the system and so
the information flow of the shared information is high. That strong driving
from the first map drives down much if any cooperative flow of synergistic
information flow, though there is a moderate, if small, amount of intrinsic
information flow.

Finally, \Cref{fig:tentmapslopetwo} reveals a large increase in
cooperative---that is synergistic---information flow when the second map (at
$a_2 \in [0, 1.3]$) is only moderately unstable. The intrinsic and shared
information flows vanish as the synergistic information peaks. Again, the
informational interpretation is that the two maps together are creating and
communicating information. Interestingly, the shared information is maximized
at $a_2 \approx 1.65$ exactly where the intrinsic and synergistic flows vanish.

Overall, one concludes that distinct information flows between coupled chaotic
systems exhibit a range of different kinds of information flow, dominated at
one or another nonlinear parameter and coupling strength by intrinsic, shared,
and synergistic flows.

As a final note, the informational estimates here are empirical. However, we
would like to point out for the case of nonlinear dynamical systems
with good measurements (obtained via their generating partitions) that
information measures such as those introduced here for flows can be
analyzed exactly. Examples of the analytical methods we have in mind
are given in Ref. \cite{Jame13a}.

\subsection{Leader-Follower Dynamics}
\label{subsec:leaderfollower}

The framework put forth here has also been applied to the study of flocking
behavior in bacteria~\cite{sattari2022modes}. There, through extensive use in a
number of different flock behaviors, it was found that the intrinsic
information flow better distinguishes the roles of leaders and followers in
collective motion. A variety of information flow types, similar to those just
seen, appear and govern to a large extent the cooperative and competitive group
behaviors.

\subsection{Financial Information Flows}
\label{subsec:financial}

We next analyze information flows between a financial index and its constituent
stocks. The value of a financial index is the weighted average of the value of
its constituent stocks. Here, we highlight our analysis of the Standard \&
Poor's 500 (S\&P 500), while
\cref{app:financial} compares information flows in the S\&P 400 and S\&P
600 indices. The S\&P 500 consists of 500 ``large cap'' stocks, whose total
value is approximately $\$23.9$ trillion dollars or 80\% of the US market. The
other indices are smaller, engineered to reflect financial dynamics at other
economic scales. The time series consist of whether the daily closing price of
each stock and of the index rose ($1$) or did not rise ($0$) between January
\nth{1} 2000 and December \nth{31} 2008. We only include the 252 stocks whose
symbol was in the index for the entirety of the date range and for which a
complete price history is available.\footnote{Index membership from
\url{https://github.com/fja05680/sp500}; prices from Yahoo Finance and, for
since-delisted stocks \url{https://www.tiingo.com}.} We estimate each
information flow measure utilizing a past of length $1$. These methods closely
follow those of Ref.~\cite{kwon2012asymmetric}, where the transfer entropy
between the S\&P 500 and its constituents was analyzed. As noted, interpreting
the transfer entropy as an information flow is unclear.

The flows are small---most well below \SI{0.005}{\bit}---and plug-in estimates
from $N = 2261$ observations are biased upward by roughly \SIrange{3 \times
10^{-4}}{6 \times 10^{-4}}{\bit} even between independent series. We therefore
test each flow against surrogate data. Each mode vanishes under particular
conditional independencies among \Xpast, \Ypres, and \Ypast: intrinsic flow
whenever $\TE$ or $\TDMI$ does, synergistic flow whenever, e.g., $\Ypres \ind
\Ypast \mathrel{|} \Xpast$. We realize each such condition by permuting one
variable within strata of another. This preserves every other pairwise
dependence, in particular the strong same-day co-movement of a stock and its
index. (See App. \ref{app:significance} for details.)

A mode is declared nonzero only if every condition under which it vanishes is
rejected, an intersection--union test~\cite{berger1982multiparameter}. Since
$\ITE \leq \min(\TE, \TDMI)$, the observed intrinsic flow is compared with the
surrogate distributions of $\TE$ and $\TDMI$. Intrinsic flow is thus
significant only when transfer entropy is---the hybrid approach of
\cref{subsec:calculation}---which avoids the excess false positives of
Ref.~\cite{de2022inferring}. We use $10^4$ surrogates per test and control the
false discovery rate at 5\% within each mode and
direction~\cite{benjamini1995controlling}. Claims about the population of
stocks are assessed with a stationary block bootstrap over trading
days~\cite{politis1994stationary}, applied jointly to all series (mean block
length 20 days, $10^3$ replicates). Details are in \cref{app:significance}.

\begin{figure}
  \centering
  \includegraphics{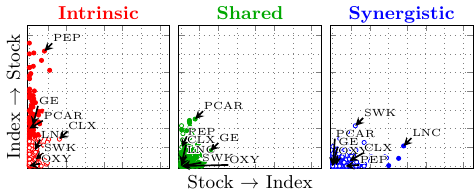}
\caption{Distinct information flows within the ``large cap'' S\&P 500 financial
	index. Each point is one of 252 stocks. Filled markers denote flows that are
	statistically significant (false discovery rate 5\%) in at least one direction;
	hollow markers, in neither. No stock has a significant flow of the same mode in
	both directions. Axis scales are identical with minimum \SI{0.0000}{\bit} and
	maximum \SI{0.0126}{\bit}.
	}
\label{fig:sp500}
\end{figure}

To probe market behavior with our more refined scheme, we evaluate the
intrinsic, shared, and synergistic information flows between each stock and its
index. The S\&P 500's information flows are given in \cref{fig:sp500}. The
analysis immediately reveals that intrinsic information flow is heavily skewed:
the index intrinsically drives 80 of the 252 stocks, while no stock has a
statistically significant intrinsic flow to the index. Across stocks,
index-to-stock intrinsic flow exceeds stock-to-index by a median of \SI{1.25
\times 10^{-3}}{\bit} (95\% confidence interval \SIrange{0.39 \times
10^{-3}}{2.34 \times 10^{-3}}{\bit}). Shared flow shows no significant
aggregate asymmetry (median difference \SI{-0.05 \times 10^{-3}}{\bit},
interval \SIrange{-0.51 \times 10^{-3}}{0.00}{\bit}). Yet it is significant
almost exclusively from stock to index: for 134 stocks, versus 2 in the
opposite direction. This reflects the index's own memory---its daily moves were
mean-reverting over this period, $\I{\Ypres : \Ypast} = \SI{5.7 \times
10^{-3}}{\bit}$ ($p \leq 10^{-4}$)---together with each stock's same-day
co-movement with the index.

A stock's past is predictive of the index's present through what it shares with
the index's past. Many of these shared flows are nonetheless small, so here
significance reflects consistency rather than magnitude. Synergistic flow is
skewed in the other direction (median difference \SI{-0.29 \times
10^{-3}}{\bit}, interval \SIrange{-0.71 \times 10^{-3}}{-0.02 \times
10^{-3}}{\bit}): some joint feature of the prior values of the stock and index
is predictive of the index value, but not so of the stock. Individually,
however, synergistic flow is significant for only five stocks, all
stock-to-index, so this asymmetry is a population-level effect assembled from
many individually undetectable contributions. In
Ref.~\cite{kwon2012asymmetric}, the asymmetry of the transfer entropy---the sum
of intrinsic and synergistic flows---in the S\&P 500 was noted. Our flow
decomposition refines this asymmetry. The excess of index-to-stock transfer
entropy is carried by intrinsic flow and is detectable stock by stock. The
smaller stock-to-index excess is synergistic and is detectable only in
aggregate.

We can further analyze the behavior of specific stocks. PEP (PepsiCo), for
example, is strongly driven---via intrinsic flow---by the behavior of the S\&P
500, as are GE (General Electric) and PCAR (PACCAR Inc). Neither GE's nor PCAR's
sizable shared flow from the index is significant. Instead, both have significant
shared flow toward it; GE's is the largest of any stock. LNC (Lincoln Financial
Group) has the largest synergistic flow of any stock, and it runs from stock to
index: LNC's past behavior combined with that of the S\&P 500 is predictive of
the index. OXY (Occidental Petroleum) operates virtually independently from the
S\&P 500, with exceedingly little information flow of any mode in either
direction. Finally, CLX (Clorox) has the largest stock-to-index intrinsic flow of
any stock, yet it is not significant: apparent intrinsic influence of individual
stocks on the index is consistent with estimation noise.

The S\&P 400 and S\&P 600 have also been analyzed, though without significance
testing, and the results are in \cref{app:financial}. These mid- and small-cap
stocks generally have less total information flow in both the stock-to-index and
index-to-stock directions. Intrinsic flow is broader in these indices, without
the pronounced asymmetry seen in the S\&P 500. Shared flow, however, shows a strong asymmetry in the
S\&P 400 while is nonexistent in the S\&P 600. Synergistic flow, again, shows
none of the asymmetry that is seen in the S\&P 500.

\subsection{Calculation}
\label{subsec:calculation}

Regarding to practical application of modes of information flow, two
angles have been explored in works published after this manuscript was made
public. Reference~\cite{zhang2019itene} explores a method of estimating the intrinsic transfer entropy in high-cardinality data. This methodology leverages
large neural network estimators in order to avoid explicitly optimizing to find
the mapping $\Ypast \to \overline{\Ypast}$. Reference~\cite{de2022inferring} 
demonstrates that if one performs statistical hypothesis testing, for a
fixed amount of data the null hypothesis is more likely to be erroneously
rejected when using the intrinsic transfer entropy than using standard transfer
entropy. This may limit the measure's usefulness when inferring the presence or
absence of intrinsic dependencies between systems, but does not imply that the
quantification of that dependence is inappropriate. Rather, it suggests a hybrid
approach: Use transfer entropy when testing for the presence of information
flow, and use the intrinsic transfer entropy to quantify its strength.
See also App. \ref{app:significance} for estimating statistical significance.

\section{Conclusion}
\label{sec:Conclusion}


Detecting and quantifying information flow is both important and
ill-defined---proposals to date have led to ambiguous, misleading, or
inconsistent interpretations of behavior and structure. Conceptually,
information flow is the medium through which causality propagates. Here, we
proposed that one of the primary impediments to successfully diagnosing
information flow is that it is not a singular concept. Rather, information flow
can take on several qualitatively distinct modes. The intrinsic information
flow is the mode most closely aligned with prior intuitions, such as that
motivating the transfer entropy~\cite{schreiber2000measuring}.

To quantify the intrinsic information flow, we proposed the cryptographic flow
ansatz that posits intrinsic flow is synonymous with the ability to construct a
secret key. This obviated the infinite regress of conditioning on all of a
system's other, possibly unspecified, components and the effects arising from
conflating conditional independence and conditional dependence. This enabled us
to quantify intrinsic information flow using the intrinsic mutual information,
an easily computed upper bound on the secret key agreement rate. With this in
hand, the remaining flow modes are quantified with the aid of the time delayed
mutual information and the transfer entropy.

When appealing to cryptographic secret key agreement rate, we made a choice to
approximate it with the intrinsic mutual information. Though, tighter upper
bounds on the secret key agreement rate exist, they are generally much more
difficult to estimate~\cite{gohari2017achieving}. This makes them generally
impractical in all but the smallest and simplest of cases. There also exist
lower bounds on the secret key agreement rate~\cite{gohari2017achieving},
though these too are computationally prohibitive for general practice.
Presumably, using improved bounds would be justified by an application's need
for more accuracy.

Refinements aside, the distinct quantification of each mode of information flow
is broadly applicable. Demonstrating its consistency and discriminating power,
we computed the intrinsic, shared, and synergistic information flows for key
base cases and between several financial indices and their constituent stocks.
As a new lens into stock market dynamics, these led to a significantly more
nuanced view of the interactions between individual companies and the market.
For example, we found that stocks whose transfer entropy from the S\&P 500 is
large are that way due to intrinsic flow, statistically significant for nearly a
third of its constituents, while no stock detectably drives the S\&P 500
intrinsically.

Additionally, shared information flow is often entirely neglected in analyses
due to the prevailing opinion that transfer entropy supplants time-delayed
mutual information whereas when considering information flow as multimodal the
latter plays a first-class role. Without observations such as these it is
impossible to paint a complete picture of how information is shuttled
throughout a complex system. Looking forward, we believe that quantifying the
distinct modes of information flow in an even broader variety of settings will
lead to substantial improvements in our understanding how a system and its
components interact to generate truly complex behavior.

\section*{Acknowledgments}
\label{sec:acknowledgments}


We thank D. Feldspar for helpful discussions. As an External Faculty
member, JPC thanks the Santa Fe Institute and the Telluride Science Research
Center for their hospitality during visits. This material is based upon work
supported by, or in part by, John Templeton Foundation grant 52095,
Foundational Questions Institute grant FQXi-RFP-1609, the U.S. Army Research
Laboratory and the U. S. Army Research Office under contract W911NF-13-1-0390
and grant W911NF-18-1-0028, and via Intel Corporation support of CSC as an
Intel Parallel Computing Center. BZ was supported by the 2017 NSF Research
Experience for Undergraduates Program and BDMA by a UC MEXUS-CONACYT Doctoral
Fellowship.


\appendix

\section{Background: Times Series and Information}
\label{app:Back}

Let us summarize notation. We denote random variables using capital letters
($X$), realizations of random variables using lower case ($x$), and the event
space of a random variable using calligraphics ($\mathcal{X}$). We denote a
sequence of temporally-ordered random variables (a time series) using a
Python-like slice notation: $X_{t}, X_{t+1}, X_{t+2}, \ldots, X_{t+\tau-1} =
X_{t:t+\tau}$. We suppress the starting or ending index of a slice if it is
infinite; to wit, a bi-infinite time series is simply $X_{:}$. Throughout we
assume stationarity---$X_{t:t+\tau} = X_{0:\tau}$---and that time series are
ergodic. A time series is IID when it's random variables are independent and
identically distributed.

We next review several fundamental information-theoretic measures; for a more
detailed introduction please refer to any standard text; e.g.,
Refs.~\cite{Cove06a,mackay2003information,Yeung2008}. The central measure of
information theory is a random variable's \emph{entropy}:
\begin{align}
  \H{X} = -\sum_{x \in \mathcal{X}} p(x) \log_2{p(x)}
  ~.
\end{align}
The entropy of a joint variable is defined similarly:
\begin{align}
  \H{X,Y} = -\sum_{\mathclap{x, y \in \mathcal{X}\times\mathcal{Y}}} p(x, y) \log_2{p(x, y)}
  ~.
\end{align}
These quantify the total amount of uncertainty that exists within a set of
random variables. Given two random variables, the \emph{conditional entropy}
quantifies uncertainty of one given knowledge of the other:
\begin{align}
  \H{X|Y} = \H{X,Y} - \H{Y}
  ~.
\end{align}
That is, it is the uncertainty ``left over'' after the uncertainty of $Y$ is
removed from the joint uncertainty of $X$ and $Y$.

Entropy is, generally, subadditive.
Its degree of subadditivity is known as the \emph{mutual information} and quantifies the dependence between two variables:
\begin{align}
  \I{X:Y} &= \H{Y} + \H{X} - \H{X,Y} \\
          &= \H{X,Y} - \H{X|Y} - \H{Y|X} \\
          &= \sum_{\mathclap{x, y \in \mathcal{X}, \mathcal{Y}}} p(x, y) \log_2{\frac{p(x, y)}{p(x)p(y)}}
  ~.
\end{align}

The \emph{conditional mutual information} measures the additional change in
uncertainty about $Y$ given $X$, when given a third variable $Z$:
\begin{align}
  \I{X:Y|Z} &= \H{Y|Z} - \H{Y|X,Z} \\
            &= \sum_{\mathclap{x, y, z \in \mathcal{X}, \mathcal{Y}, \mathcal{Z}}} p(x,y|z) \log_2{\frac{p(x,y|z)}{p(x|z)p(y|z)}}
  ~.
\end{align}
Note that conditioning can increase statistical dependence; that is, $\I{X:Y|Z}
> \I{X:Y}$. This reflects the fact that conditional mutual information is
sensitive to both intrinsic dependencies between $X$ and $Y$, as well as
dependencies induced by $Z$. In other words, dependence between $X$ and $Y$ may
be revealed through their relationship with $Z$. Such dependencies can occur
even when $X$ and $Y$ are independent: $\I{X:Y} = 0$.

References~\cite{Crut01a,Jame11a} review how these elementary information
quantities extend to measure randomness and correlation in time series.

\section{Joint Analysis of Information Flow}
\label{app:joint}

To gain a greater understanding of how joint interactions among time series
lead to the three modes of information flow, consider all variables of interest
simultaneously. \Cref{fig:idiagram} represents all interactions of \Xpast,
\Xpres, \Ypast, and $\overline{\Ypast}$ in the form of an
\emph{I-diagram}~\cite{Yeun91a}. Three of the regions are identically zero due
to the constraints placed on $\overline{\Ypast}$; namely, that the variables
form a Markov chain $\Xpast\Ypres$---$\Ypast$---$\overline{\Ypast}$. The time
delayed mutual information $\TDMI = a + b + c$; the transfer entropy $\TE = a$;
and the intrinsic flow $\ITE = a + b$.
Since the intrinsic mutual information is bound from above by the conditional
mutual information, we conclude that information atom $b$ is necessarily
nonpositive.

As noted, intrinsic flow is given by $a + b$. Shared flow is then given by the
time delayed mutual information minus the intrinsic flow, which is $c$.
Synergistic flow is given the transfer entropy minus the intrinsic flow,
that is, $-b$. Together, these measure the total flow or $a + c$.

\begin{figure}
  \centering
  \begin{tikzpicture}[every node/.style={transform shape}, scale=1.0]
    \idiagramfour

    \draw [very thick] \W;
    \draw [very thick] \X;
    \draw [very thick] \Y;
    \draw [very thick] \Z;

    \node at (Wlabel) [anchor=east] {\Xpast};
    \node at (Xlabel) [anchor=east] {\Ypres};
    \node at (Ylabel) [anchor=west] {\Ypast};
    \node at (Zlabel) [anchor=west] {$\overline{\Ypast}$};

    \node at (WZgXY) {0};
    \node at (WXZgY) {0};
    \node at (XZgWY) {0};
    \node at (WXgYZ) {$a$};
    \node at (WXYgZ) {$b$};
    \node at (WXYZ)  {$c$};
  \end{tikzpicture}
  \caption{
    Analysis of how the three measures of information flow are related.
    From the definition of intrinsic mutual information, $\I{\Xpast \Ypres : \overline{\Ypast} | \Ypast} = \I{\Xpast : \overline{\Ypast} | \Ypast} = \I{\Ypres : \overline{\Ypast} | \Ypast} = 0$.
    Furthermore, $b \leq 0$. $\TDMI = a + b + c$, $\TE = a$, and $\ITE = a + b$.
    From this, we can determine that intrinsic information flow is $a + b$, shared information flow is $c$, and synergistic flow is $-b$.
    Together, the total information flow is $a + c = \I{\Xpast : \Ypres} - \I{\Xpast : \Ypres : \Ypast | \overline{\Ypast}}$.
  }
  \label{fig:idiagram}
\end{figure}

\section{Relationship With the Partial Information Decomposition}
\label{app:pid}

The separation of transfer entropy into two components is not a new idea.
Utilizing the \emph{partial information decomposition}~\cite{williams2010nonnegative} Williams \etal~\cite{williams2011generalized} decompose the transfer entropy from $X$ to $Y$ into two components: that \emph{uniquely} provided by \Xpast (``state independent transfer''), and that \emph{synergistically} provided by both \Xpast and \Ypast (``state dependent transfer''):
\begin{align}
  \TE =& \I[\cap]{\Xpast \rightarrow \Ypres \setminus \Ypast} + \\
       & \I[\cap]{\Xpast \Ypast \rightarrow \Ypres}
  ~.
\end{align}
While our intrinsic and synergistic information flows are qualitatively very similar to this idea, there are some important distinctions.
First, it is well-known that the intrinsic mutual information, used here to quantify intrinsic flow, is incompatible as a measure of unique information within the partial information decomposition~\cite{bertschinger2013shared}.
Second, the goal in the partial information decomposition is to decompose (for example) $\I{\Xpast, \Ypast : \Ypres}$.
Here, we do not constrain our flows to any particular sum.
This is in line with the lack of any conservation law governing information~\cite{james2016information}.
Furthermore, with the lack of widely accepted method of computing the partial information decomposition and the hints of incompleteness when the number of inputs exceeds two~\cite{rauh2017secret}, there is always the unfortunate possibility that such an endeavor simply can not be realized.

\section{Computing the Intrinsic Mutual Information}
\label{app:computing}

All computations were performed with \texttt{dit}~\cite{dit}. Its
implementation of the intrinsic mutual information utilizes the basin hopping
method from \texttt{scipy.optimize}. While this calculation is straightforward
for discrete probability distributions, there are significant difficulties in
computing the intrinsic mutual information for continuous random variables. In
essence, and barring a particularly clever and unforeseen method, one must
establish a repertoire of potential transformations to apply to the variable
$Z$. The conditional mutual information then must be minimized over the space
of these transformations. Its evaluation can be performed using estimation
methods such as the standard KSG estimator~\cite{kraskov2004estimating}.

\begin{figure}
  \centering
  \includegraphics{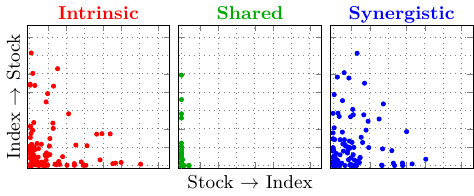}
\caption{Information flows within the ``mid cap'' S\&P 400.
	Axis scales are identical with minimum \SI{0.0000}{\bit} and
	maximum \SI{0.0025}{\bit}.
  }
  \label{fig:sp400}
\end{figure}

\begin{figure}
  \centering
  \includegraphics{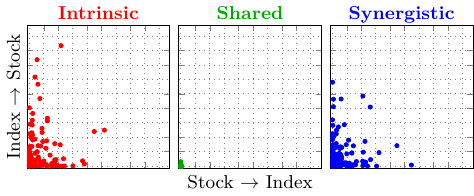}
\caption{Information flows within the ``small cap'' S\&P 600.
	Axis scales identical with minimum \SI{0.0000}{\bit} and
	maximum \SI{0.0032}{\bit}.
  }
  \label{fig:sp600}
\end{figure}

\section{Financial Indices Detailed Analysis}
\label{app:financial}

Here, we compare our analysis of the S\&P 500 financial index in the main text
to flows detected in the S\&P 400 and S\&P 600 indices. The S\&P 400 consists
of 400 ``mid cap'' stocks, whose total value is approximately $\$2.1$ trillion
dollars or 7\% of the US market. The S\&P 600 consists of 600 ``small cap''
stocks, whose total value is approximately $\$896$ billion dollars or 3\% of
the US market.

Again, the time series consist of the sign of the change in the daily closing price of each stock and the index between January \nth{1} 2000 and December \nth{31} 2008 for stocks whose symbol was in the index for the entirety of the date range. And, as for the S\&P 500, we estimate each information flow measure utilizing a past of length $1$.

\subsubsection{S\&P 400}
\label{subsubsec:sp400}

The information flows within the S\&P 400 are illustrated in \cref{fig:sp400}.
These mid cap stocks display very different dynamics than the large cap S\&P
500. Neither the intrinsic nor the synergistic flows display any marked
asymmetry in directionality. The shared information flow, however, demonstrates
a strong asymmetry where the stock and index are both predictive of the stock.
Overall, the amount of information flow is significantly smaller than in the
S\&P 500.

\begin{figure}
  \centering
  \includegraphics{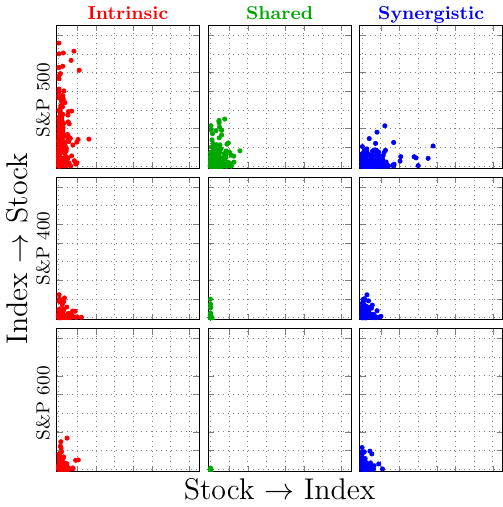}
\caption{Information flows in the S\&P 500, 400, and 600 stock indices compared.
	Axis scales identical with minimum \SI{0.0000}{\bit} and
	maximum \SI{0.0126}{\bit}.
  }
\label{fig:spall}
\end{figure}

\subsubsection{S\&P 600}
\label{subsubsec:sp600}

The S\&P 600 dynamics are again different from the large and mid cap stocks.
The intrinsic flows are skewed in a similar manner as the S\&P 500, though not
as strongly. Shared information flow, however, is effectively nonexistent.
Synergistic flows are largely symmetric, similar to the mid cap stocks.

The dynamics of all three indices are displayed on equal scales in \cref{fig:spall}.

\section{Statistical Significance of Financial Flows}
\label{app:significance}

The flow measures for financial time series are a concern as they are so low.
The concern is compounded as the flow measures themselves are all nonnegative
quantities and so empirical estimates are biased positive. And this increases
the likelihood of specious positive estimates, when there are no flows.

So, to refine our tests of statistical significance of informational estimates
we define \emph{permuting one variable within strata of another} to mean
holding the value of one variable fixed and shuffling others. This is
a particularly helpful test for financial times series given our observation
that the informational estimates are so low.

For example, to permute $X_{-1}$ within the \emph{strata} of $Y_{-1}$, we take
the entire time series of $X_{-1}, Y_{-1}, Y_0,$ and identify the rows where
$Y_{-1} = 0$ (call this subset $A$) and where $Y_{-1} = 1$ (call that subset
$B$). Then, take all the $X_{-1}$ values in subset $A$ and randomly permute
them, keeping their distribution the same, but shuffling relative to $Y_0$.
Then do the same within subset $B$. This makes $X_{-1}$ conditionally
independent of $Y_0$ given Y$_{-1}$. In this way, whatever the values estimated
for the transfer entropy, intrinsic flow, or synergistic flow from $X$ to $Y$
they will be within the ``noise'' distribution since, in theory, the shuffling
has rendered them independent. In short, whatever the estimated quantity---e.g.,
the transfer entropy---for the new partially permuted time series, that value
is purely spurious.

With this mind, write $X = \Xpast$, $Y = \Ypres$, and $Z = \Ypast$.
\Cref{tab:surrogates} lists four surrogate families. Each permutes one
variable, globally or within strata of another, and so imposes exactly one
conditional independence while preserving the remaining pairwise marginals.

\begin{table*}
  \centering
  \begin{tabular}{lll}
    \toprule
    Permute & Imposes & Consequence \\
    \midrule
    $X$ within strata of $Z$ \hspace{0.15in} & $X \ind Y \mathrel{|} Z$
	\hspace{0.15in} & $\I{X : Y \mid Z} = \I{X : Y \downarrow Z} = 0$, so synergistic flow $= 0$ \\
    $X$ globally             & $X \ind (Y, Z)$          & $\I{X : Y} = \I{X : Y \mid Z} = 0$ \\
    $Z$ within strata of $X$ & $Y \ind Z \mathrel{|} X$ & synergistic flow $= 0$, shared flow $= \I{Y : Z}$ \\
    $Z$ within strata of $Y$ & $X \ind Z \mathrel{|} Y$ & synergistic flow $= 0$, shared flow $= \I{X : Z}$ \\
    \bottomrule
  \end{tabular}
  \caption{Surrogate families used to test each flow mode, with $X = \Xpast$, $Y = \Ypres$, and $Z = \Ypast$.}
  \label{tab:surrogates}
\end{table*}

The last two rows follow from
$\I{X : Y \mid \overline{Z}} = \I{X : Y} - \I{Y : \overline{Z}}$
when $Y \ind Z \mathrel{|} X$, which is minimized by $\overline{Z} = Z$, and
symmetrically for $X \ind Z \mathrel{|} Y$. For each mode, and each family $f$ under which
it vanishes, we form $p_f = (1 + \#\{T_b \geq T_{\text{obs}}\}) / (1 + B)$ with
$B = 10^4$ surrogates, and report $p = \max_f p_f$. For synergistic flow, $T$ is the
synergistic flow itself under the first, third, and fourth families. For intrinsic
flow, $T_{\text{obs}}$ is the observed $\I{X : Y \downarrow Z}$ and $T_b$ is the
surrogate $\I{X : Y \mid Z}$ (first family) or $\I{X : Y}$ (second family). This is
valid since $\I{X : Y \downarrow Z}$ is bounded by both for every sample.

Shared flow vanishes when $Y \ind Z$ or $X \ind Z$. The third and fourth families
retain the dependence mediated by the remaining variable, so their surrogate shared
flows are stochastically larger than under those conditions, and the test is
conservative. Two limitations remain.
\begin{itemize}
  \item The intrinsic test covers the sub-hypotheses $\I{X : Y \mid Z} = 0$ and
  $\I{X : Y} = 0$, but not distributions in which $\I{X : Y \downarrow Z}$ vanishes
  only through a nontrivial channel $Z \to \overline{Z}$.
  \item Permutation treats the overlapping observation triples as exchangeable.
\end{itemize}
Temporal dependence is addressed for the population-level claims by the stationary
block bootstrap.

\bibliography{chaos,ref,ite}

\end{document}